\documentclass[pdflatex,sn-mathphys-num,iicol]{sn-jnl}
\usepackage{graphicx}  
\usepackage{url}                  
\usepackage{dcolumn}       
\usepackage{bm}                  
\usepackage{amssymb}       
\usepackage{mathtools}     
\DeclarePairedDelimiter\ket{\lvert}{\rangle}
\usepackage{amsmath}    
\usepackage{pgf}                   
\usepackage{eurosym}
\usepackage{tikz,pgfplots}                   
\usetikzlibrary{shapes,arrows,calc}
\usetikzlibrary{arrows, decorations.markings}
\usetikzlibrary{decorations.pathreplacing,calligraphy}
\usepackage{xcolor}
\usepackage{listings}
\usepackage{quantikz}

\begin{document}  

\title[The potential of quantum computers for Particle Image Velocimetry]{The potential of quantum computers for Particle Image Velocimetry}

\author*[1]{\fnm{Philipp} \sur{Pfeffer}}\email{philipp.pfeffer@tu-ilmenau.de}

\author[1,2]{\fnm{Theo} \sur{Käufer}}\email{thka2078@mit.edu}

\author[1]{\fnm{Julia} \sur{Ingelmann}}\email{julia.ingelmann@tu-ilmenau.de}

\author[1]{\fnm{Christian} \sur{Cierpka}}\email{christian.cierpka@tu-ilmenau.de}

\author[1]{\fnm{Jörg} \sur{Schumacher}}\email{joerg.schumacher@tu-ilmenau.de}

\affil[1]{\orgdiv{Institute of Thermodynamics and Fluid Mechanics}, \orgname{Technische Universität Ilmenau}, \orgaddress{\street{P.O.Box 100565}, \city{Ilmenau}, \postcode{98684}, \country{Germanyy}}}

\affil[2]{\orgdiv{Department of Mechanical Engineering}, \orgname{Massachusetts Institute of Technology}, \orgaddress{\city{Cambridge}, \postcode{02139}, \state{MA}, \country{USA}}}

\date{\today}

\abstract{
Particle Image Velocimetry (PIV) is the prime image-processing technique to measure and visualize velocity fields of laminar and turbulent flows. The velocity field vectors  are obtained with sub-pixel accuracy by analyzing cross-correlations, empowered by Fast Fourier Transforms (FFT). Here, we present a quantum algorithm with multidimensional quantum Fourier Transforms, termed Quantum-based PIV (QuPIV), to replace the classical computation of up to millions of velocity vectors. Our end-to-end quantum algorithm includes a novel state preparation, modified amplitude amplification, and the output extraction. We enhance amplitude amplification by a contracted ground-state projector, which allows a significant reduction of the number of gates in the quantum circuit. We justify the end-to-end capability with numerical studies on all stages of the algorithm on both synthetic and experimental data.}

\maketitle

\section{Introduction}
The cross-correlation and convolution stand among the most common and frequently used operations in signal processing~\cite{smith1997scientist}. They allow to filter~\cite{byerly1965convolution}, compare~\cite{Raffel2018,Westerweel}, and extract~\cite{lecun1998gradient} features from data and are irreplaceable  in science and industry. In particular, the cross-correlation is an ubiquitous building block in today's artificial intelligence~\cite{lecun1998gradient}, astronomy data analysis~\cite{hayashi2008understanding}, weather forecasting~\cite{zuo2024improved}, medical image processing~\cite{sarvaiya2009image}, fluid mechanics~\cite{Raffel2018}, and beyond~\cite{farmer1997review}. With this omnipresence in mind, refining these algorithms would pave the way for an increased efficiency for a wide range of real-world data processing applications.

\begin{figure*}
\centering
\includegraphics[width=0.94\linewidth]{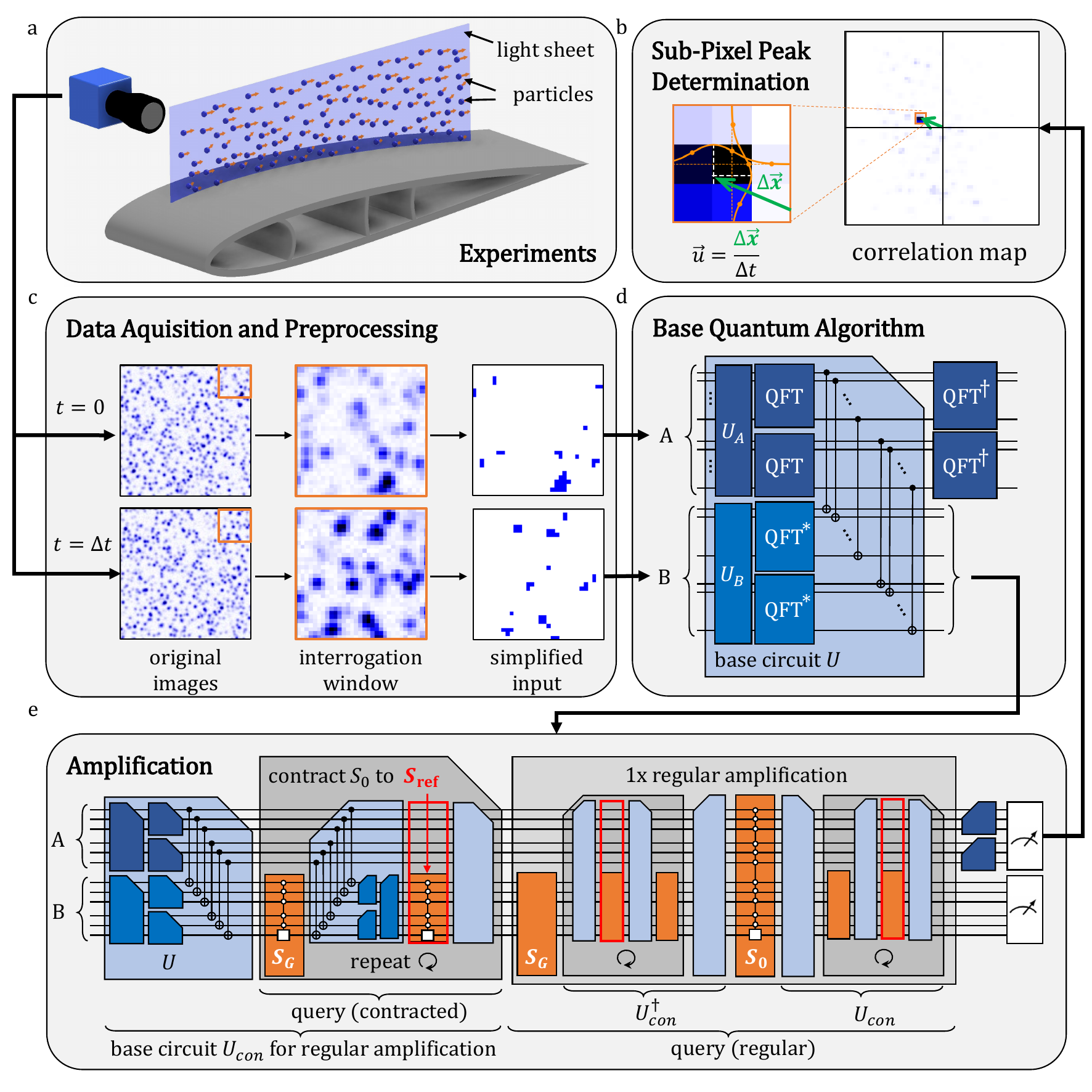}
\caption{Building blocks and workflow of Quantum-based Particle Image Velocimetry.  (a) The input data for the algorithm, which are obtained from PIV experiments, are tracer particle images. Here, they are obtained for a flow around a wing. (b) Evaluation of the peak of the correlation map $C_{\rm map}$, which determines the velocity field vector in each local interrogation window to sub-pixel accuracy. (c) Preprocessing of PIV snapshots by filtering to binary input images. For an $N\times N$ interrogation window, the $N$ pixels with the largest light intensities are taken. (d) Image loading on separate qubit registers $A$ and $B$ followed by the Fourier-based cross-correlation evaluation. We apply a two-dimensional Fourier Transform of the images with the multidimensional quantum Fourier transformation (QFT). For the complex conjugated operation QFT$^*$, we conjugate every gate which then constructs the complex conjugated spectrum on register $B$ required for $C_{\rm map}$ which is given by Eq. \eqref{eq:ClCorr}. The component-wise product is already given by parallel preparation; thus we only have to sort it by the central block of controlled NOT quantum gates. This sorting assures that the component-wise product elements occur in those cases, where the overall quantum state is in basis state $\ket{0}$ on all qubits in the lower register  $B$. (e) We modify amplitude amplification by contraction of the ground state projector $S_0$ and insert this into one regular amplitude amplification. The last inverse QFTs realize the back-transformation of the component-wise product and we finally access the correlation map via almost every measurement. The peak position is easy to measure as it aggregates the largest amplitudes, corresponding to the highest measurement probabilities, see again (b). Qubits in (d,e) start in basic state $\ket{0}$. The arrows between the panels indicate the workflow.}   
\label{fig:QuPIV}
\end{figure*}

A prominent application of cross-correlation for fluid mechanics is Particle Image Velocimetry (PIV) to determine velocity vector fields~\cite{kahler2016main}. It is know as the standard technique for this task in science and is wide-spread in industry due to the high achievable resolution~\cite{ChristianJ.Kahler.2012b,Scharnowski_ruleofthumb_2020} and its robustness~\cite{Raffel2018}. The centerpiece of the PIV algorithm are many cross-correlations that determine the shift or displacement of tracer particles between two digital images. Small tracer particles are chosen to follow the flow nearly perfectly. Once properly distributed in the fluid, they are illuminated by a laser light sheet and then captured by one or more cameras, as illustrated in Fig.~\ref{fig:QuPIV} (top left). The particle image displacement in an interrogation window thus quantifies the fluid motion statistically~\cite{Raffel2018,Westerweel}. 

The cross-correlation produces the correlation map, which is the spatial distribution of correlation coefficients for pixel-wise input displacements. The correlation of a given displacement between two inputs $A$ and $B$ is the amplitude on a pixel that is displaced equally  from the image center. The largest correlation values rise around the actual mean particle image displacement, which is called the correlation peak~\cite{Westerweel}. Its characteristic shape even enables sub-pixel accurate peak position determination ($\sim 0.1 \ldots 0.05$ pixel), as illustrated in panel (b) of Fig.~\ref{fig:QuPIV}. The displacement value is the actual output of the algorithm. The velocity is obtained by dividing the displacement vector by the image time lag. The velocity in physical coordinates is traced by pixel scaling or advanced calibration schemes.

The computational efficiency of PIV stems from the Fast Fourier Transform (FFT). The correlation map is computed by fast correlation methods that adapt fast convolution algorithms based on the convolution theorem. The actual computational challenge of PIV is that high spatial and temporal resolution requires a very large number of correlation evaluations. This accumulates for advanced methods that help to retrieve reliable vector fields in the case of low signal-to-noise-ratio (SNR) or high velocity range. Prominent examples are the sum-of-correlation approach~\cite{Westerweel2004}, single pixel ensemble correlation~\cite{Scharnowski2011} or the pyramid cross-correlation~\cite{Sciacchitano2012}. From this perspective, even modest speedups of the base algorithm would be invaluable in view of experimental studies for turbulent flows. 

It is known for the fast convolution algorithm that the Quantum Fourier Transform (QFT) can be used for a one-dimensional quantum convolution algorithm~\cite{Ramezani23}. However, the fast correlation requires complex conjugation and in the case of PIV multidimensional QFTs~\cite{Multi_QFT}. Both steps can be implemented efficiently, as outlined later. The special use-case of PIV also allows to address known problems for end-to-end quantum algorithms~\cite{PRA_Comment}. At input level, the essential information are particle positions, not the full image. At output level, the essential information is the correlation peak position, not the complete correlation map. This can be used to reduce the complexity of state preparation and sampling drastically, making the use of quantum computers promising. 

Here, we present an end-to-end quantum algorithm for the signal displacement estimation, which is the major ingredient of PIV. Our algorithm pipeline includes a novel state preparation scheme, the two-dimensional quantum cross-correlation evaluation, a valuable adaption of amplitude amplification together with an expressive statistical analysis of measurement and processing errors. Especially the adaption of amplitude amplification also provides relevant insights for many other quantum algorithms where amplitude amplification is necessary. We analyze connections between the different algorithmic stages and verify these by simulations of a typical PIV analysis based on synthetic and experimental input. Thereby, our work demonstrates the use of quantum algorithms for PIV and sets a new impetus in quantum computing by discussing an industrial relevant application on end-to-end level.

As quantum computing differs considerably from classical comuputing a brief introduction is given in Section~\ref{sec:Intro}. In Section~\ref{sec:Algorithm Setup} the algorithm setup of QuPIV is described in general, including the state preparation~(\ref{subsec:State Preparation}), the 2D cross-correlation~(\ref{subsec:2DCC}) and different amplitude amplification schemes~(\ref{subsec:Standard Amplification}, \ref{subsec:Contracted Amplification}). Section~\ref{sec:Results} compares the results between QuPIV and classical PIV based on real experimental images. The key findings are subsequently summarized and discussed in Section~\ref{sec:discussion}.

\section{A primer on quantum computing}\label{sec:Intro}
We briefly summarize some basic definitions of quantum computing, see also Nielsen and Chuang for more details~\cite{Nielsen2010}. While a single classical bit can take two discrete values, namely $\{0,1\}$ only, a single quantum bit (in short qubit) is a superposition of two basis states in the vector space $\mathbb{C}^2$ which can take any state on the surface of a unit sphere
\begin{equation}
|q_1\rangle=c_1|0\rangle+c_2|1\rangle=c_1
\left(
\begin{array}{c}
1 \\ 0\\
\end{array}\right)
+c_2
\left(
\begin{array}{c}
0 \\ 1\\
\end{array}\right)\,, \label{Eq:B1}
\end{equation}
with $c_1, c_2\in \mathbb{C}$ and $\sqrt{|c_1|^2+|c_2|^2}=1$ . Vectors $|0\rangle$ and $|1\rangle$ are the basis vectors in Dirac notation~\cite{Nielsen2010}. A two-qubit state vector is the tensor product of two single-qubit vectors, 
\begin{equation}
|q_2\rangle=|q_1\rangle\otimes |q_1^{\prime}\rangle\,.
\end{equation}
The basis of this tensor product space is given by 4 vectors: $|j_1\rangle=|0\rangle\otimes |0\rangle$, $|j_2\rangle=|0\rangle\otimes |1\rangle$, $|j_3\rangle=|1\rangle\otimes |0\rangle$, and $|j_4\rangle=|1\rangle\otimes |1\rangle$. An $n$-qubit quantum state, which is given by
\begin{equation}
   \ket{\Psi} = \sum\limits_{k=1}^{2^n} c_k \ket{j_k} \hspace*{1em} \text{with} \hspace*{1em} \sum\limits_{k=1}^{2^n} |c_k|^2 = 1\,,
\end{equation}
is called {\em fully separable} if it can be written as
\begin{equation}
   \ket{\Psi} = \overset{n}{\underset{i=1}{\bigotimes}} \,\ket{q_i} \,,
\end{equation}
where $\ket{q_i}$ are single qubit quantum states given by Eq.~\eqref{Eq:B1}. It is called {\em separable} if a tensor product decomposition of $\ket{\Psi}$ into blocks is possible with at least one multi-qubit quantum state $\ket{q_i}$, that is not fully separable. Not separable multi-qubit quantum states are called {\em entangled}. An $n$-qubit quantum state is called {\em fully entangled} if no subspace of separable qubits exists. 

Quantum computing corresponds to an imposed time evolution of a single or multiple quantum states (typically denoted as a quantum register). In a closed quantum system, this evolution of the quantum register is linear and reversible, thus realized by unitary transformations $U$ such that
\begin{equation}
\ket{\Psi(t)} = U(t) \ket{\Psi(0)} \quad \mbox{with} \quad U^{-1}(t)=U^{\dagger}(t)\,.
\label{eq:uni1}
\end{equation}
This unitary time evolution is realized by the Schrödinger equation,
\begin{equation}
i\hbar\frac{\partial \ket{\Psi(t)}}{\partial t} = H\ket{\Psi(t)}\,.
\label{eq:uni2}
\end{equation}
The Hamiltonian operator $H=T+V$ is self-adjoint, i.e. $H=H^{\dagger}$, and represents the kinetic and potential energy of the quantum system. The Schrödinger equation~\eqref{eq:uni2} can be solved formally giving 
\begin{equation}
\ket{\Psi(t)} = \exp\left(-\frac{i}{\hbar}Ht\right) \ket{\Psi(0)}\,.
\label{eq:uni3}
\end{equation}
Compared with Eq.~\eqref{eq:uni1}, we immediately see that the matrix exponential in Eq.~\eqref{eq:uni3} satisfies $U(-t)=U^{-1}=U^{\dagger}$ since the Hamilton operator is self-adjoint. An isolated quantum system is a strong idealization, which is hard to reach in reality even though the devices such as superconducting quantum computers are typically operated at millikelvin temperatures.~\cite{Krantz2019}

A quantum algorithm is composed of a quantum circuit that acts on $n$ qubits (the quantum register) to perform a specific numerical operation. The algorithm consists of a sequence of single- and two-qubit gates along each qubit. Gates are elementary unitary transformations. Quantum parallelism comes as a consequence that the unitary operator $U$ acts on all components of the quantum state at the same time in parallel,
\begin{equation}
U \ket{\Psi} = \sum_{k=0}^{2^n-1} U [c_k \ket{k}]\,.
\label{eq:uni4}
\end{equation}
The readout of information is done by measurement of observables. The latter is a strong intervention into the isolated quantum system that leads to a spontaneous collapse of the quantum state onto one of the eigenstates of the observable. The measurement result is thus always of probabilistic nature.

\section{Algorithm Setup}\label{sec:Algorithm Setup}

The full workflow of the QuPIV algorithm is shown in Fig.~\ref{fig:QuPIV}. It starts with the fluid dynamical experiment applying PIV using a laser light sheet, e.g., around a wing as shown in panel (a) and ends with the final determination of the velocity $\vec{u}$ in panel (b). This originates from the correlation map $C_\text{map}$ of two PIV images at successive times $t=0$ and $t=\Delta t$. Reduced and binarized as illustrated in the {\em Data Aquisition and Preprocessing} block in panel (c), they are loaded to the $2n$-qubit quantum registers A and B before being processed. The {\em Base Quantum Algorithm} block in panel (d) depicts the quantum version of the fast correlation algorithm, given by 
\begin{equation}
    C_{\text{map}} = \mathcal{F}_{2D}^{-1} \left\{ \mathcal{F}_{2D}\{A\} \odot \mathcal{F}^\star_{2D}\{B\} \right\}.
    \label{eq:ClCorr}
\end{equation} 
$\mathcal{F}_{2D}$ denotes the two-dimensional Fourier Transform, complex conjugation is indicated by '$^\ast$' and '$\odot$' is the component-wise or Hadamard product. The Hadamard product is a non-unitary step and demands amplitude amplification, which we modify in the {\em Amplification} block as illustrated in panel (e). As amplitude amplification uses for- and backward circuit executions, we depict the direction with a cut-out corner on the upper right (forward) or left (backward) in panels (d) and (e). The measurement of the amplified cross-correlation brings us back to panel (b), where the velocity vector $\vec{u}$ is obtained by the peak shift $\Delta \vec{x}$ from the map center and the time difference $\Delta t$ between the subsequent images, in short $\vec{u}=\Delta\vec{x}/\Delta t$. 

The central challenge for many end-to-end quantum algorithms is the communication bottleneck between classical and quantum computers, see e.g. Zhang et al. (2021)~\cite{Zhang2021} and Nakaji et al. (2022)~\cite{Nakaji2022}. This can be called the quantum von-Neumann bottleneck. Quantum states offer exponentially growing resources under efficient processing, namely the state elements under quantum parallelism. However, preparing exponentially many amplitudes is often as difficult as extracting exponentially many amplitudes via sampling~\cite{Aaronson2015,Hashim2025}. Therefore, we put strong focus on state preparation and amplitude amplification in the following, while the actual sampling of the correlation map is efficient and outlined in the Supplementary Note "Sampling" together with Supplementary Fig. 3.

We provide a statistical error analysis on all stages of the algorithm in the Supplementary Notes "Image reduction", "Amplitude amplification" and "Sampling" together with Supplementary Figs. 1, 2, 3. This focuses on the peak position, which is the relevant output of the algorithm. We identify the sub-pixel position of a correlation peak by standard three-point interpolation~\cite{Nobach2005}, as indicated in panel (b) of Fig.\ref{fig:QuPIV}. Error sources can shift the original peak position $\Delta \vec{x}_0=(x_0,y_0)$ to a different one $\Delta \vec{x}_1=(x_1,y_1)$. The standard error measure of this work is the shortest distance between those peaks, namely the euclidean norm
\begin{equation}
    E = \|\Delta\vec{x}_0-\Delta\vec{x}_1\|_2 = \sqrt{(x_0-x_1)^2+(y_0-y_1)^2}\,.
    \label{eq:peak_error}
\end{equation}
This error needs a statistical evaluation to be expressive for a given setup. We generate 500 different pairs of square-sized particle image distributions with a fixed displacement of $N/4$ in both dimensions for an edge length of $N=32$, $64$, $128$ and $256$ pixels. The qubit number $n$ thus formally corresponds to the image edge length $N=2^n$. However, the standard for classical correlation is not the circular cross-correlation that assumes periodic boundary conditions, but the linear cross-correlation. This doubles the edge size by zero-padding and results in $(n+1)$ qubits per edge to obtain unbiased results. We use the circular cross-correlation in the main text because it is sufficient for the experimental data we evaluate. On the synthetic data in the Supplementary Material, we use linear cross-correlation such that both approaches are studied within this work. For simplicity, we illustrate the $n$ qubit per edge notation in the following to be consistent with the circular correlation throughout the main text. We use the bra-ket notation for (multi-qubit) quantum states~\cite{Nielsen2010}. We abbreviate the controlled $X$ or 'NOT' gate as cNOT and the same gate but with multiple controllers as mcNOT. Also, the little-endian convention is used where the least significant bit occurs at the end of the bitstring and the upper end of the circuit. 

\begin{figure*}
    \centering
    \includegraphics[width=1\linewidth, trim={2cm 0 2cm 0},clip]{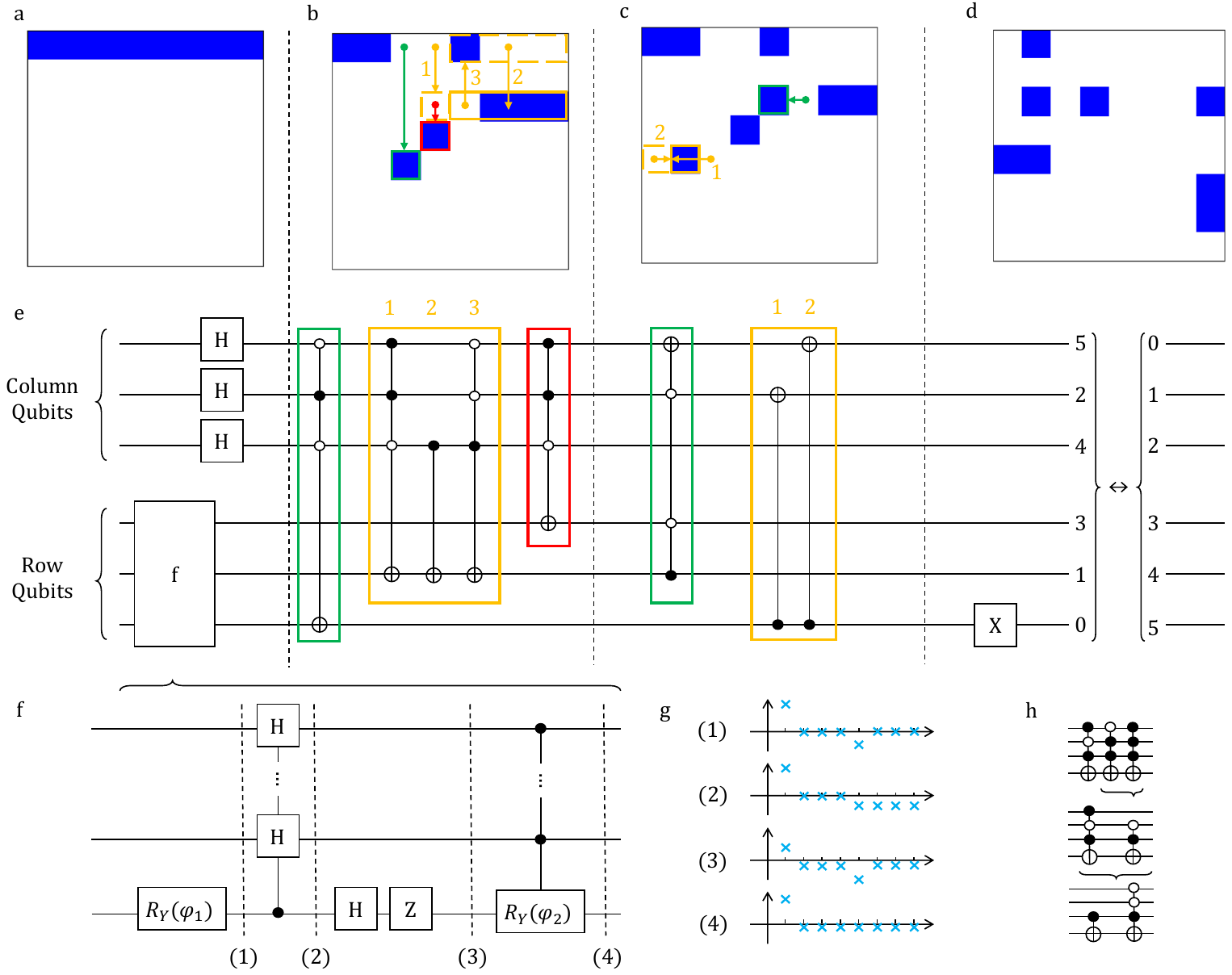}
    \caption{Efficient quantum state preparation for binary images with zero mean. The procedure is illustrated for clarity by $n=3$ qubits in both registers for the $N=8$ brightest pixels in an $8\times 8$ pixel image. (a)-(d) Stepwise iteration to the final image and corresponding quantum circuit blocks in panel (e). Panel (e) shows the gates for the example target image in panel (d). The individual pixel shifts and the corresponding multi-controlled NOT (mcNOT) gates are boxed in the same color as in panels (b) and (c). Panel (a) shows the initial image, which has a zero mean. (f) Details of the circuit diagram for a binary and zero-mean state on the row qubits in panel (e). (g) Illustration of the stages (1)-(4) in panel (f). The gate arguments $\varphi_1$ and $\varphi_2$ depend on the image size and are given in the main text. All qubits in panel (e) and (f) are supposed to start in state $\ket{0}$. (h) An example of the standard controller simplification as used within the stage of (b) in panel (e) by aggregating mcNOT gates is illustrated.}
    \label{fig:State_Prep}
\end{figure*}

\subsection{Quantum state preparation}\label{subsec:State Preparation}
We design a specific version of sparse state preparation~\cite{Ramacciotti2024,Mao2024}. While the quantum circuit for a sparse state can be comparably shallow, their classical advantage is that identifying this circuit can also be highly efficient as only a limited amount of elements needs to be tracked. The first step is to find a sparse representation of the input. We use the $N$ pixels with the largest intensities of an $N\times N$ image and set them to uniform intensity. This data reduction is acceptable, as discussed in Supplementary Note "Image Reduction". It allows to identify the sparse representation efficiently because only the $N$-th largest element needs to be identified. This can be done by modifying the routine median of medians~\cite{Blum1973}. For the particle images in this work, we can even exploit that high intensities are comparably rare in contrast to pixels representing the irrelevant background. This motivates an iterative thresholding, which we tested as follows. 

We identify the minimum and maximum pixel amplitudes on the complete image and take their average. We select all elements (or pixels) that are above the average and track the minimum of these elements. For our synthetically generated images, such a step always gives more than $N$ out of the total $N^2$ elements. We repeat this first iteration with the previous average as the new minimum. If less than $N$ elements are above the new average in a subsequent iteration, we reduce the maximum to the average and track how many elements are still required. Otherwise, the minimum is lifted to the average as before. This iteration proceeds until less than we have identified the largest $N$ values. On our test data, we always find the $N$ largest elements with less than $4N^2$ comparisons. As we compare our algorithm with classical FFTs, we emphasize that both the latter as well as many of the following operations are comparisons of bits. These usually require simple boolean operations, whereas FFTs operate with floating point operations.

With the $N$ positions of largest intensities identified, we discard their amplitudes and focus on their location only. This provides more flexibility in state preparation, since all relevant elements are now equal and can thus be permuted without practical impact. Under these conditions, we start our state preparation with a default initial image where all $N$ pixels are in the first row, as illustrated in panel (a) of Fig.~\ref{fig:State_Prep}. Such an initial configuration can be realized with $n$ Hadamard gates on the $n$ column qubits and the $\ket{0}$ state on all $n$ row qubits. However, we find that mean-value-free images show better sampling capabilities because a mean on the inputs also produces a mean on the output. This corresponds to a relevant probability to measure anywhere in the correlation map. To prevent this, we prepare the row qubits in a binary, but zero-mean or mean-value-free state as illustrated in panel (f) and (g). 

A binary and mean-value-free state is obtained by one large positive amplitude $a$ and many small negative amplitudes $b$ for the $N$ elements on the row qubits. The Hadamard gates on the column qubits then duplicate this first column of structure $(a,-b,-b,\dots,-b)$ to the full image, which retains a zero mean. The amplitudes follow by
$$ a=\sqrt{\frac{N-1}{N}} \quad\mbox{and} \quad  b=\frac{1}{\sqrt{N(N-1)}}.$$
The circuit, that produces this vector in stages (1)--(4) is shown in panel (f) of Fig.~\ref{fig:State_Prep}. The relevant angles in the $R_Y$ rotation gates of the circuit follow by
$$ \varphi_1 = 2\sin^{-1}\left[ \frac{-1}{\sqrt{N-1}} \right]$$ 
and 
$$  \varphi_2= 2\cos^{-1} \left[\frac{N-1}{\sqrt{N(N-2)+2}}\right]\ ...$$ \\
$$+2\tan^{-1}\left[ \frac{-\sqrt{2}}{\sqrt{N(N-2)}} \right]. $$
Recall that $R_Y(\varphi)=e^{-iY\varphi/2}$ with the Pauli gate $Y$. 

The remaining circuit stages in panel (e) of Fig.~\ref{fig:State_Prep} are pixel shifts as exemplified with panels (b) to (d). We identify these stages backwards, which means that we start from the target image (d) and reduce it to the default initial image of panel (a). This is motivated by summarizing the most challenging operations in stage (c), while the following stage (d) is highly efficient. This last stage just organizes row and column qubits. The position of a pixel in terms of the quantum state is given by its row and column bitstring. A mapping between bits and qubits formally imposes no operations while the flexibility can align the target and the initial image. 

We map the qubits in stage (d) by counting how often each of the $2n$ bits is in state $|0\rangle$ or $|1\rangle$ over the complete list of $N$ pixels. This yields localized bits that are predominantly in one state, and balanced bits which are approximately equally distributed between both states. The initial image is fully localized in the row qubits and fully balanced in the column qubits. Thus, we rearrange all bitstrings such that localized bits are the first bits (row qubits) and balanced bits are the last bits (column qubits). Both types are sorted by their criterion, so the first row qubit is the most localized bit. If a row qubit is localized towards the $1$ state, we flip it globally by a NOT gate, which corresponds to a Pauli gate $X$ in the quantum circuit, for instance as shown in panel (e) of Fig.~\ref{fig:State_Prep}.

From stage (c) to (b) in Fig.~\ref{fig:State_Prep}, all relevant pixels become unique by their column position. This corresponds to one pixel per column and allows the remaining operations in stage (b) to be controlled by the column qubits alone. However, these shifts are complex to realize without potentially shifting other elements. The detailed methodology for this step is given in the Supplementary Note "State preparation". To summarize, we identify columns with multiple relevant pixels and move all except one element into an empty column each. For these shifts, we prioritize pixels that are in rows without any other relevant pixel because there the row position suffices to avoid other shifts while realizing column shifts.

After each pixel has been assigned to its own column, stage (b) uses the column qubits as controllers to shift all pixels into their rows. We start with the most significant row qubit, as illustrated in panel (e), and search all pixels where this qubit is in state $\ket{1}$. This results in a list of bitstrings that correspond to control sequences on the column qubits. While these could all be executed individually, we can aggregate these sequences to use both less and simpler gates, as illustrated in panel (h) of Fig~\ref{fig:State_Prep}. For instance, the strings $00$ and $01$ can be summarized as $0x$, using only one gate with one controller. The detailed reduction rules are outlined in the Supplementary Note "State preparation".

This concludes the image preparation, which is realized in the circuits $U_A$ for the image at $t=0$ and $U_B$ for $t=\Delta t$ in the base quantum algorithm, see panel (d) of Fig.~\ref{fig:QuPIV}. For linear cross-correlation, we add one ancilla qubit for zero-padding as a new most significant bit per dimension. A cNOT gate controlled from the former most significant bit to the ancilla followed by an $X$ gate on the controller can shift the image into the center of the padded image.
 
\subsection{Two-dimensional cross-correlation}\label{subsec:2DCC}
Adapting the convolution theorem by complex conjugation of one spectrum allows to realize cross-correlation similar to an existing (quantum) convolution method~\cite{Ramezani23}. The resulting correlation map $C_\text{map}$ is already given by Eq.~\eqref{eq:ClCorr}. For this adaption, we need a two-dimensional QFT, see ~\cite{Multi_QFT}, and the complex conjugation. We use the fundamental property $(U_1U_2\ket{\Psi})^\star=U_1^\star U_2^\star\ket{\psi}^\star$ for arbitrary unitaries $U_1,U_2$ and state $\ket{\psi}$. As the input data is real-valued, this only affects the two QFTs. Then QFT$^\star$ has the same circuit as the QFT, but with conjugate arguments in the phase gates. In contrast, an inverse QFT also inverses all other gates and  reverses the sequence of gate applications.

The central challenge in adapting the classical method is the component-wise product, which is a nonlinear step. An accessible way to realize this product is to prepare both images and thereby both spectra on separate qubit registers. We denote these registers as $A$ for the image at $t=0$ and $B$ for $t=\Delta t$.  This encodes not only the component-wise product, but literally every pairwise product between the two spectra. The component-wise product for any spectral mode pair $\hat{A}_k\hat{B}_k$ is encoded at the basis state $\ket{k}_B\otimes\ket{k}_A$ of the global state. The necessary back-transformation requires to align these elements, such that an inverse QFT can be used. This can be done with a cNOT from each qubit in one register as controller, here $A$, to each respective qubit in the other register~\cite{Holmes2023}, as illustrated in Fig.~\ref{fig:QuPIV}(d). This transforms the relevant elements from $\ket{k}_B\otimes\ket{k}_A$ to $\ket{0}_B\otimes\ket{k}_A$. Thus all mode pairs are properly sorted and a two-dimensional inverse QFT on the upper register $A$ concludes Eq.~\eqref{eq:ClCorr}. However, the correlation map is only provided if register $B$ is in state $\ket{0}_B$.

\subsection{Standard amplitude amplification}\label{subsec:Standard Amplification}
Amplitude amplification is a common routine to increase the amplitude of target elements while reducing irrelevant state elements~\cite{Brassard1997,Grover1996,Ramezani23}. This uses a base unitary matrix $U$ that prepares a state denoted with $\ket{\Psi_U}$. The corresponding $U$ is depicted in Fig~\ref{fig:QuPIV}(d) with light-blue background. We can decompose $\ket{\Psi_U}$ by
\begin{equation}
    \ket{\Psi_U} =\sin(\theta)\ket{\mathcal{T}}+\cos(\theta)\ket{\mathcal{T}_\perp}
    \label{eq:Ampl_Sep}
\end{equation}
where $\ket{\mathcal{T}}$ contains target elements and $\ket{\mathcal{T}_\perp}$ the orthogonal complement. Consider $\ket{\mathcal{T}}$ as the first $N^2$ elements of $\ket{\Psi_U}$ in normalized form and thus $0$ everywhere else. These are the elements of the spectral correlation map $C_{\rm map}$ since $\ket{0}_B$ holds. To map $\ket{\mathcal{T}}$ and $\ket{\Psi_U}$ in Eq.~\eqref{eq:Ampl_Sep}, the normalization of $\ket{\mathcal{T}}$ introduces a prefactor which defines the base amplitude $a_0=\sin(\theta)$ with base angle $\theta$. The probability to measure target elements is called {\em success probability}. For $r$ amplification steps, which are termed {\em queries}, the success probability is $p_r=a_r^2$ where $a_r$ is the success amplitude given by
\begin{equation}
a_r = \sin((2r+1)\theta). 
\label{eq:Ampl_Amp}
\end{equation}
A query stands for the execution of $US_0U^\dagger S_G$. The diagonal unitaries $S_0$ and $S_G$ are $S_0=\text{diag}\{-1,1,...,1\}$ and in this work $S_G=\text{diag}\{-1,...,-1,1...,1\}$ with the first $N^2$ elements at -1. One realizes 
\begin{equation}
S_0=X^{\otimes 4n}Z_{4n}X^{\otimes 4n}
\end{equation}
with $Z_q=\text{diag}\{1,,...,1,-1\}$ as a multi-controlled $Z$ gate over all $q=4n$ qubits in the circuit. $S_G$ is constructed equally, but only  $q=2n$ qubits of register $B$ are necessary, as shown in Fig.~\ref{fig:QuPIV}(e). Due to this locality of $S_G$, we can reduce the base circuit as shown in Fig.~\ref{fig:QuPIV}(d). The terminating QFT$^\dagger$ operations only require register $A$. Considering $U^\dagger S_G U$ as a sequence produced with amplification queries, the terminating QFT$^\dagger$ in the base quantum algorithm is not separated by $S_G$ from the initial QFT in $U^\dagger$, and thereby QFT$^\dagger\,$QFT$=\mathbb{I}$. Only the last $U$ has no partner to reduce QFT$^\dagger$, so this QFT$^\dagger$ remains as shown at the end of Fig. \ref{fig:QuPIV}(e). 

For the optimal query count $r_\text{opt}$ where $a_{r_\text{opt}}\approx 1$, the base angle $\theta$ or amplitude $a_0$ needs to be known. Though $a_0$ can be approximated by sampling the base circuit, we show in the Supplementary Note "Amplitude amplification" that for a given image size and type, the mean of the corresponding success probability is approximately one order of magnitude larger than its standard deviation (std). In practice, where many comparable images are processed, the overhead to identify $r_\text{opt}$ can thus be strongly mitigated because the first identified $r_\text{opt}$ is also a suitable candidate for subsequent images. We also show that the initial success probability $p_0$ scales with $1/N^2$, which means that $r_\text{opt}$ scales with $N$~\cite{Ramezani23}.

\subsection{Contracted amplitude amplification}\label{subsec:Contracted Amplification}
Considering amplification queries as repeated executions of $US_0U^\dagger S_G$, further circuit reductions between $U$ and $U^\dagger$ would be possible if operators $S_0$ or $S_G$ act on fewer qubits. We denote this strategy of reducing the circuit with an adapted $S_0$ or $S_G$ as {\em contracted amplitude amplification}. We do not contract $S_G$ in this work due to the complexity, which will be further outlined in the Discussion section. 

Let us assume that we replace $S_0$ with $S_\text{ref}$, which marks the ground state on $m$ qubits in a circuit with $q>m$ qubits and define $\kappa=q-m$. We also assume that $S_G$ acts on the same $m$ qubits as $S_\text{ref}$, while potentially using fewer than those $m$ qubits. The standard amplitude amplification can then act as if $2^{\kappa}$ separate states are provided. This can be expected if the $\kappa$ and $m$ qubits are not entangled, but we find it also to hold if the amplification can be decoupled between the $\kappa$ and $m$ qubits. The concept of decoupling is explained in detail in the Supplementary Note "Amplitude amplification". In brief, assume that only a set of terminating cNOT gates controlled by the $\kappa$ qubits entangle the $\kappa$ with the $m$ qubits. In analogy to Eq.~\eqref{eq:Ampl_Sep}, this allows to separate the global state $\ket{\Psi_U}$ into $2^\kappa$ local targets $\ket{\mathcal{T}_k}$ and corresponding complements $\ket{\mathcal{T}_{k,\perp}}$ as
\begin{equation}
    \ket{\Psi_U} = \sum\limits_{k=0}^{2^{\kappa}-1} c_k \left[ \sin(\theta_k)\ket{\mathcal{T}_k}+\cos(\theta_k)\ket{\mathcal{T}_{k,\perp}} \right]\,.
    \label{eq:Ampl_Contr}
\end{equation}
The relevant elements of $\ket{\mathcal{T}_{k}}$ and $\ket{\mathcal{T}_{k,\perp}}$ are the rescaled elements of $\ket{\Psi_U}$ where the $\kappa$ qubits are in state $\ket{k}$. The elements that are also mapped by $S_G$ are in $\ket{\mathcal{T}_{k}}$. Both vectors are normalized to map the regular notation of amplitude amplification. As a consequence, we introduce an additional prefactor $c_k$ that accounts for the combined norm of $\ket{\mathcal{T}_{k}}$ and $\ket{\mathcal{T}_{k,\perp}}$ in $\ket{\Psi_U}$. Also, $\theta_k$ now defines the local base angle of $\ket{\mathcal{T}_{k}}$ in respect to $\ket{\mathcal{T}_{k,\perp}}$. With $r$ amplification queries, a target element grows by $\sin((2r+1)\theta_k)$, similar to Eq.~\eqref{eq:Ampl_Amp}. As $\theta_k$ can vary with $k$, the contracted amplification modifies the relation between target elements. This error has to be analyzed in context of potential circuit reduction. 

We now focus on our use-case. If $S_\text{ref}$ avoids either register $A$ or $B$, then one full image preparation with subsequent QFTs is not required in $U^\dagger$ and $U$ with $US_\text{ref}U^\dagger$. We avoid register $A$ here, so $S_\text{ref}$ and $S_G$ both span the register $B$, as depicted in Fig.~\ref{fig:QuPIV}(e). This also simplifies the identification of $c_k$ and $\theta_k$. As the register $A$ corresponds to the previous $\kappa$ qubits, $\ket{\mathcal{T}_k}$ and $\ket{\mathcal{T}_{k,\perp}}$ are all elements where we find $\ket{k}_A$. While the central cNOTs in the base algorithm entangle register $A$ with register $B$, the permutation of state elements occurs in register $B$, so the state on register $A$ remains unaffected. This means that for any given $k$, the parameters are
\begin{equation}
    c_k = |\hat{A}_k| \hspace{2em} \sin(\theta_k) =\frac{|\hat{A}_k\hat{B}_k|}{|\hat{A}_k|}=|\hat{B}_k|
    \label{Eq:SimpAA}
\end{equation}
because $c_k$ is the norm of all states with $\ket{k}_A$. As these are all elements of the normalized spectrum on register $B$ multiplied with $\hat{A}_k$, the norm of all these elements is $| \hat{A}_k|$. The corresponding  $\sin(\theta_k)$ is given by the amplitude of the target element $|\hat{A}_k\hat{B}_k|$ divided by the maximum amplitude $c_k$. In conclusion, simulating the effect of contracted amplification for our application task can be efficiently described by the spectra of the input images $\hat{A}$ and $\hat{B}$. 

The errors of contracted amplitude amplification can be balanced by error-free regular amplitude amplification. In Fig.~\ref{fig:QuPIV} we show the setup with a single regular amplification query. We amplify with contracted amplification towards the amplitude of $0.5$ and then use this circuit as a base circuit for the standard amplitude amplification. This requires a single query for unit success probability. To verify this step, insert $r=1$ into Eq.~\eqref{eq:Ampl_Amp} with $\sin^{-1}(0.5)$ as $\theta$; this gives $\sin(3\sin^{-1}(0.5))=1$. While further standard amplifications queries and thus less contracted amplification queries reduce the error further, we show that this is not necessary, both for synthetic data in the Supplementary Material and experimental data in the following section.

\section{Evaluation with experimental data}\label{sec:Results}
\begin{figure*}
    \centering
    \includegraphics[width=0.97\linewidth]{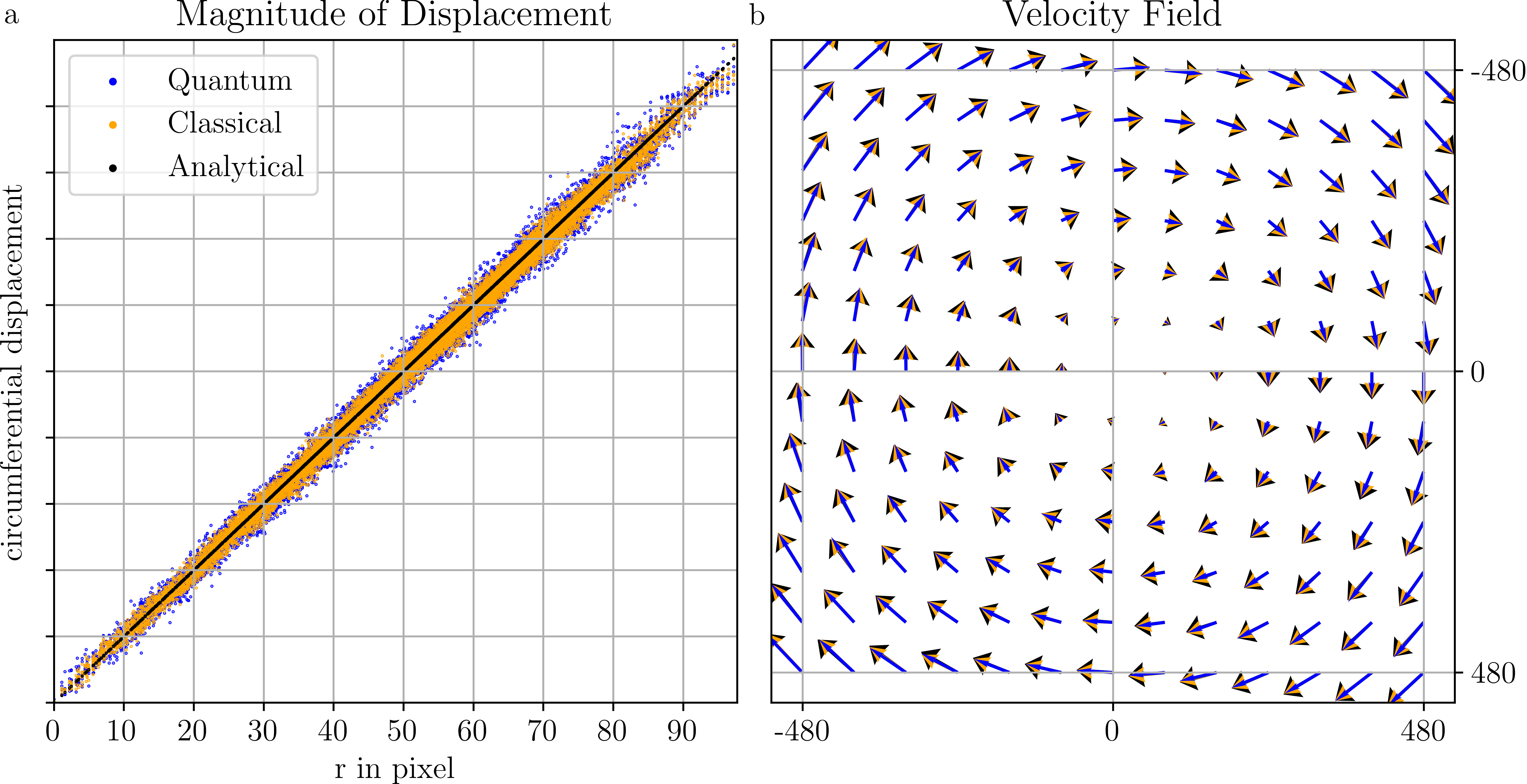}
    \caption{Evaluation of the 4th PIV challenge data (case F) with the QuPIV pipeline via a comparison of the analytical solution (black), classical PIV with binary images (orange) and the full QuPIV pipeline with additional contracted amplification and sampling. (a) All $121^2$ displacement amplitudes based on their radial position. The circumferential displacement is normalized such that it aligns with the radial pixel displacement from the image center. (b) Velocity vector field comparison where every 10th vector is depicted based on its distance from the image center in pixel. Parameters are given in the main text.} 
    \label{fig:3}
\end{figure*}
To evaluate the performance on true experimental data, we choose case F of the 4th PIV challenge as a benchmark~\cite{kahler2016main}.
The case consists of two frames of a liquid column rotating at a constant rate. The benefit of this approach is that the ground truth of the cases is precisely known. We compare our quantum algorithm with the solution obtained using classical cross-correlation-based techniques of the OpenPIV software package~\cite{liberzon2021openpiv}. The data is analyzed on the same grid as required for the PIV challenge and presented in Fig.~\ref{fig:3}. To compare the analytical (black) with the quantum (blue) and classical (orange) solution, we use $64\times 64$ pixel interrogation windows with 87.5$\%$ overlap and perform only a single circular correlation pass. In both cases, we use binary images with $64$ active pixels. This procedure was chosen to compare the errors due to the corresponding evaluation pipeline and not due to the lower information content of the binary images.  
In the quantum case, these are the processing error due to contracted amplification and the sampling error due to finite measurement of the quantum state. 

We simulate 15 contracted amplification steps for all image pairs. This is the minimal amount to amplify every case as close as possible to $0.5$, and to this end a worst case if no further knowledge is given on individual image pairs. We sample each correlation map 500 times and choose this comparably large number to depict the general reliability of the approach without further post-processing. In detail, $99.9\%$ of all peak position errors remain below 1 pixel with this parameterization. Also, we observe that larger peak position errors could be detected without knowing the ground truth as those instance predominantly occur as isolated outliers. In Fig.~\ref{fig:3}.b, we reduce the full vector fields of $121\times121$ displacement vectors to a $13\times 13$ vector field each for illustrative purposes. Fig.~\ref{fig:3}.b illustrates that the differences between all solutions are barely visible, so we depict in Fig.~\ref{fig:3}.a the amplitude of the velocity over to the radial position in the rotating flow. There, we observe that all velocity amplitudes remain close to the analytical case and no systematic deviations from the ground truth can be seen. This demonstrates that the quantum algorithm can perform the evaluation with feasible parameters for amplification, sampling, and state preparation.

\begin{figure*}
    \centering
    \includegraphics[width=0.97\linewidth]{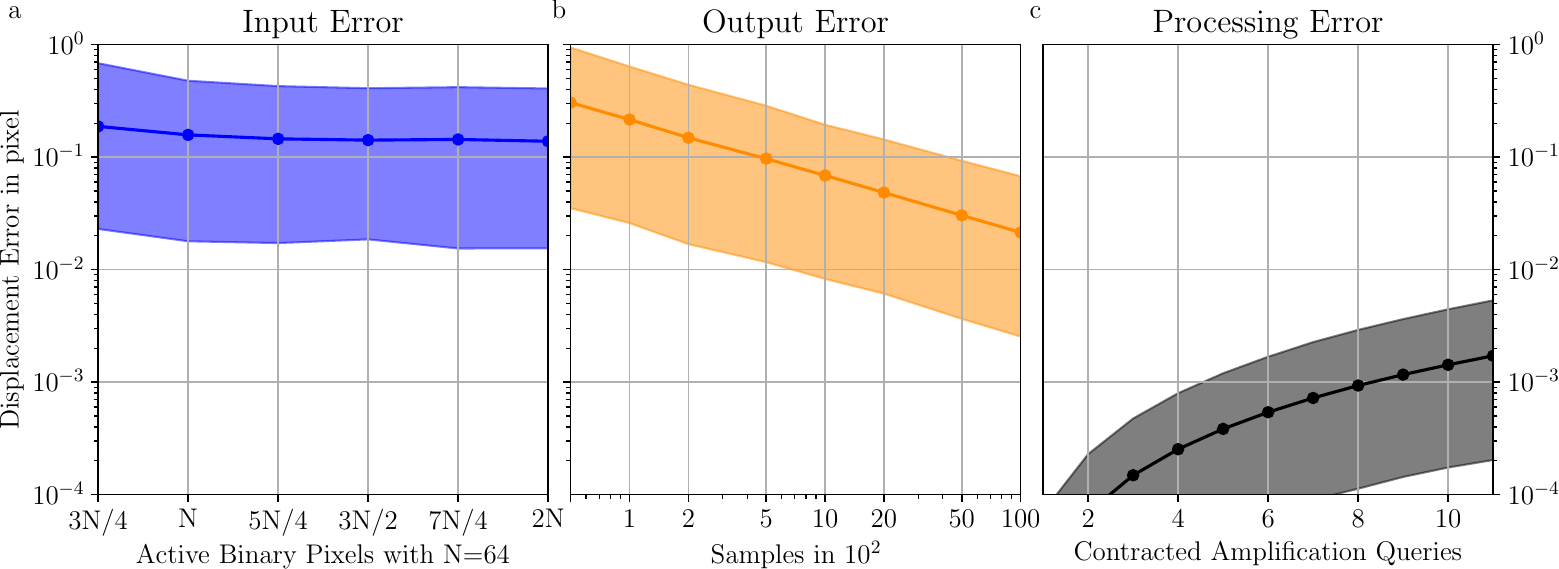}
    \caption{Quantification of error sources of the quantum algorithm with experimental data. All errors are analyzed separately. The median of each error is illustrated in a solid line and the corresponding $5\%-95\%$ percentile with a colored area. These quantities are evaluated over $121^2$ interrogation window pairs from the PIV challenge data as discussed in the main text. (a) Statistical quantities of the input error due to sparse binary images. (b) The output error due to sampling with finitely many measurements. (c) The processing error due to contracted amplitude amplification as discussed in the main text, estimated by the proposed prediction method.} 
    \label{fig:4}
\end{figure*}

For better understanding of each error source, we analyze all errors individually on this data, assuming for each test that the other errors are not present. The results are shown in Fig.~\ref{fig:4}. In comparison to default classical processing, there are three error sources which arise due to the quantum algorithm, namely an input, output and processing error.  All errors are shown by their median with a solid line and the $5\%-95\%$ percentile is shown as a correspondingly colored area. These statistical quantities are extracted over all $121^2=14641$ displacements, so numerically the $5\%$ and correspondingly $95\%$ percentile shows the $733$th best or worst peak displacement error.

The input error arises due to the sparsification of the binary input images. Depicted in Fig.~\ref{fig:4}.a, it can be seen that using more pixels beyond $N=64$ pixels only leads to marginal improvements. As an upper limit for this test, we use $2N$ active pixels because the additional information by more active pixels gets balanced with the error of uniform amplitude on these pixels. For the quantum state preparation, less active pixels are easier to prepare due to sparsity and overall necessary shifts, but using less than $N$ pixels quickly increases the error. While $3N/4=48$ active pixels show comparable statistical behavior, the loss of identifiable peaks already becomes an issue, but occurs in less than $5\%$ of all cases, so all statistical quantities are still well-defined. An evaluation with $N/2=32$ active pixels produced more unidentifiable peaks, and is thus not shown in the plot. Using $N=64$ pixels for the evaluation in Fig~\ref{fig:3} is thus motivated by the comparably stable error in connection with the proposed quantum state preparation.

The output error arises due to statistical sampling of the correlation map. Depicted in the middle frame of Fig.~\ref{fig:4}, it can be seen that all statistical quantities follow a power law that corresponds to the Monte-Carlo sampling error~\cite{Shapiro2003}, where $M$ samples produce a statistical error scaling as $1/\sqrt{M}$. Our lower limit for the number of samples in this test is $50$ because using less, for instance $20$ samples, leads to a relevant proportion of peaks that are located at completely different positions than the actual peak. To demonstrate reliability without further post-processing, we chose $500$ samples for the coupled evaluation here. However, note that many outliers due to this error could be detected by comparison with neighboring peaks~\cite{Westerweel2005}.

Lastly, the processing error due to contracted amplitude amplification is illustrated in black on the right panel of Fig.~\ref{fig:4}. Here, we use Eq.~\eqref{eq:Ampl_Contr} and Eq.~\eqref{Eq:SimpAA} as outlined in the Supplementary Material to predict the peak displacement error without the necessity to simulate deep $24$ qubit circuits. To be consistent with the other errors, we show the evaluation for the original images as inputs. For these original inputs, $11$ contracted amplification queries are sufficient to amplify all correlation map amplitudes to the target of $0.5$, as outlined in the contracted amplitude amplification section. We observe that the error for this parameter is even below $10^{-2}$ pixels for all statistical quantities. In the Supplementary Material, we repeat this evaluation for synthetic images and find that more amplification queries are needed for the simplified binary images, but the errors for binary inputs perform comparable to the original input images. For the experimental inputs in binary form, we find that $15$ amplification queries are sufficient to amplify towards the $0.5$ target for all input pairs, which is consistent with the synthetic data in the Supplementary Material, where the same parameter would be $14$. Note that the Supplementary Material uses linear cross correlation, so there the $N=32$ case results in $64\times 64$ pixel images. In conclusion, we use $15$ amplification queries for the analysis in Fig.\ref{fig:3} due to the binary images, but the statistical quantities for binary images show strong similarity to those of original images, which are depicted in Fig.~\ref{fig:4}.c.

\section{Discussion}\label{sec:discussion}
This work provides a proof-of-concept for an end-to-end quantum algorithm under experimental Particle Image Velocimetry conditions with viable advances in state preparation and amplitude amplification, the latter of which defining two essential building blocks of the method. PIV builds on numerously repeated evaluations of image correlations; it is thus transformable into a quantum processing task for a moderate number of qubits, appropriate for near-term quantum computing devices. Furthermore, the presented detailed description of contracted amplitude amplification serves as a general and accessible tool for quantum computing which reaches beyond the discussed PIV case.  We emphasize that Eq.~\eqref{eq:Ampl_Contr} allows to simulate the effects and thus errors under contracted amplitude amplification without the potentially infeasible circuit simulation. The input vector $\ket{\Psi_U}$ suffices to provide such predictions while the necessary decoupling can be verified by basic knowledge of the circuit architecture. 

Keeping these benefits in mind, we see room for improvement of our QuPIV algorithm in the state preparation which is based on sparse and homogeneous data. As shown both for experimental (see Fig.~\ref{fig:3}) as well as synthetic data (see the Supplementary Material), the image reduction is the central error source of the algorithm while also requiring more amplification queries. Simultaneously, the sparse binary input renders the Fourier-based approach irrelevant on classical devices since all non-zero correlation coefficients can be identified by analyzing all pair-wise shifts between the $N$ pixels of image A and B, resulting in $N^2$ pixel pairs. However, the general problem for any state preparation is the trade-off between circuit identification and circuit efficiency, and the herein state preparation demonstrates this in an accessible way. 

A gate-efficient state preparation for the binary input images can indeed be found, but in view of the presented strategies here, presumably at the cost of comprehensive classical pre-processing. The main question is whether the reduction of the input images towards a binary and thus pixel shift problem is the best way to address this problem, and if so, how the heuristics, that we presented here, can be further optimized. One the one hand, pixel shifts are easy to identify classically by Boolean operations with the bit-wise pixel location. On the other hand, the theoretical complexity of this permutation problem cannot surpass the classical approach with our preliminaries. 

To see this, recall that we use $N$ pixels in an $N\times N$ pixel image and repeat this state preparation approximately $N$ times with amplification queries. The limit for the state preparation is thus proportional to $N$ in comparison with a pixel-pair evaluation as described before. As we can assume, that the particle images and thereby the target positions occur randomly on an input image, any standardized placement of pixels requires corrections which scale with $N$, the amount of used pixel positions, multiplied by the length of their position bitstring,  $\log(N^2)=2\log(N)$. While corrections can be summarized to a single operation, the general correction can still require a control sequence proportional to $\log(N)$, which in turn require $2\log(N)$ Toffoli gates even if $\log(N)$ ancilla qubits are available~\cite{Nielsen2010}.

However, the goal of this work is not to show that a quantum computer can surpass the classical computer in terms of theoretical scaling, but under practical conditions for which $N$ is not arbitrary large. The images for PIV are always of limited size and higher spatial resolution requires many small interrogation windows~\cite{ChristianJ.Kahler.2012b}, as indicated in Fig.~\ref{fig:QuPIV}. Even a small constant speed-up has to be multiplied with the large number of correlation evaluations, thus leading to an advantage. To achieve this, the combination of state preparation and amplitude amplification is paramount.

Contracted amplitude amplification is the central theoretical advancement of this work since it drastically reduces the end-to-end complexity by repeating only one instead of two state preparations. This description provides practical and predictable strategies for completely different settings, such as quantum computational fluid mechanics~\cite{Zecchi2025}. While we contract $S_0$ only in this work, a contraction of $S_G$ is also possible. In our case, this would amplify irrelevant elements, which reduces the amplification rate of the actual targets. Most importantly, the potential reduction in $U^\dagger S_G U$ is not obvious as the primary target is the image preparation in $U$. As $U$ ends with the QFT, relevant reduction requires that either the QFT is also reduced or the QFT does not intercept the reduction. Both turns out to be too complex for this work, thus we do not contract $S_G$ here. However, a specifically designed state preparation ending with many local operations on either the row or column registers can motivate to contract $S_G$ towards the corresponding other register. This cancels the calls of cNOT operations and the QFTs in the base algorithm $U$, which then also allows to cancel these preparation steps within $U^\dagger S_G U$. 

To summarize the discussion, we would like to emphasize that larger input images combine several beneficial properties. We show in the Supplementary Fig. 1 that the $N\times N$ synthetic images with $N=128$ pixels or more could even use $N/2$ pixels for state preparation instead of $N$ active pixels to fall reliably below the sub-pixel peak position error of $10^{-1}$. Keeping the aforementioned problem of binary images in mind, the herein state preparation converges significantly faster if only half as many elements are necessary due to viewer shifts and increased flexibility due to sparsity. Furthermore, these images are also easier to sample. As shown in the Supplementary Note "Sampling", $M=100$ shots suffice to identify the peak position close to a peak position error of $10^{-1}$ pixel. Lastly, Supplementary Fig.3 shows that these images (with $N\geq128$) also provide consistently low peak position errors under contracted amplitude amplification. These errors even remain below $10^{-2}$ pixel when amplified to the target amplitude $0.5$. Due to these properties, we see a promising direction for quantum computing in PIV.

\section*{Acknowledgments}
P.P. and T.K. were supported by the project no. P2018-02-001 "DeepTurb -- Deep Learning in and of Turbulence" of the Carl Zeiss Foundation and the Deutsche Forschungsgemeinschaft under grant no. DFG-SPP 1881. J.I. and P.P. are also supported by the European Union (ERC, MesoComp, 101052786). Views and opinions expressed are however those of the author(s) only and do not necessarily reflect those of the European Union or the European Research Council. Neither the European Union nor the granting authority can be held responsible for them. The computations were partially carried out on Fujitsu's 40-qubit quantum computer simulator during the Quantum Simulator Challenge 2024 and we would like to thank Masayoshi Hashima, Yusuke Kimura and Takuto Komatsuki from Fujitsu Limited, Japan for their support. 

\section*{Author Contributions}
T.K. and C.C. provided the motivation and classical background of this work. P.P designed the quantum algorithm and the analysis on end-to-end capability. J.I. conducted the circuit simulations on the Fujitsu Quantum Simulator to verify the state preparation and amplitude amplification predictions. All authors analyzed and discussed the results and wrote the manuscript. 

\section*{Competing Interests}
The authors declare no conflict of interest.

\section*{Data Availability}
We will provide code and data in an open-access format once this work is accepted in a peer-reviewed journal.

\bibliography{references.bib}

@article{liberzon2021openpiv,
  title={OpenPIV/openpiv-python: OpenPIV-Python v0. 22.3},
  author={Liberzon, Alex and Lasagna, Davide and Aubert, Mathias and Bachant, Pete and Mahmoodtabar, Erfan and K{\"a}ufer, Theo and Bauer, Andreas and Vodenicharski, Boyko and Dallas, Cameron and Yang, Eric and others},
  journal={Zenodo},
  year={2021}
}

@article{Aaronson2015,
    author = "Aaronson, S.",
    title = "{Read the fine print}",
    doi = "10.1038/nphys3272",
    journal = "Nat. Phys.",
    volume = "11",
    pages = "291--293",
    year = "2015"
}

@article{Blum1973,
  title={Time bounds for selection},
  author={Blum, M. and Floyd, R. W. and Pratt, V. R. and Rivest, R. L. and Tarjan, R. E. and others},
  journal={J. Comput. Syst. Sci.},
  volume={7},
  pages={448--461},
  year={1973}
}

@Article{Westerweel2005,
  author    = {Westerweel, Jerry and Scarano, Fulvio},
  journal   = {Exp. Fluids},
  title     = {Universal outlier detection for PIV data},
  year      = {2005},
  pages     = {1096--1100},
  volume    = {39},
  doi       = {10.1007/s00348-005-0016-6},
  publisher = {Springer Science and Business Media LLC},
}

@article{Nakaji2022,
  title = {Approximate amplitude encoding in shallow parameterized quantum circuits and its application to financial market indicators},
  author = {Nakaji, K. and Uno, S. and Suzuki, Y. and Raymond, R. and Onodera, T. and Tanaka, T. and Tezuka, H. and Mitsuda, N. and Yamamoto, N.},
  journal = {Phys. Rev. Res.},
  volume = {4},
  pages = {023136},
  year = {2022},
  doi = {10.1103/PhysRevResearch.4.023136}
}

@article{ChristianJ.Kahler.2012b,
 author = {C. J. K{\"a}hler and S. Scharnowski and C. Cierpka},
 year = {2012},
 title = {On the resolution limit of digital particle image velocimetry},
 pages = {{1629--1639}},
 volume = {52},
 journal = {Exp. Fluids},
 doi = {10.1007/s00348-012-1280-x}
}

@article{Zhang2021,
  title = {Low-depth quantum state preparation},
  author = {Zhang, X.-M. and Yung, M.-H. and Yuan, X.},
  journal = {Phys. Rev. Res.},
  volume = {3},
  pages = {043200},
  year = {2021},
  doi = {10.1103/PhysRevResearch.3.043200}
}

@INPROCEEDINGS{Brassard1997,
  author={Brassard, G. and Hoyer, P.},
  booktitle={Proceedings of the Fifth Israeli Symposium on Theory of Computing and Systems}, 
  title={{An exact quantum polynomial-time algorithm for Simon's problem}}, 
  year={1997},
  volume={},
  pages={12-23},
  doi={10.1109/ISTCS.1997.595153}}

@inproceedings{Grover1996,
author = {Grover, L. K.},
title = {A fast quantum mechanical algorithm for database search},
year = {1996},
isbn = {0897917855},
publisher = {Association for Computing Machinery},
address = {New York, NY, USA},
doi = {10.1145/237814.237866},
booktitle = {Proceedings of the Twenty-Eighth Annual ACM Symposium on Theory of Computing},
pages = {212–219},
numpages = {8},
location = {Philadelphia, Pennsylvania, USA},
series = {STOC '96}
}

@article{Hashim2025,
  title={Practical introduction to benchmarking and characterization of quantum computers},
  author={Hashim, A. and Nguyen, L. B. and Goss, N. and Marinelli, B. and Naik, R. K. and Chistolini, T. and Hines, J. and Marceaux, J. P. and Kim, Y. and Gokhale, P. and others},
  journal={PRX Quantum},
  volume={6},
  pages={030202},
  year={2025},
  publisher={APS}
}

@ARTICLE{Nobach2005,
       author = {{Nobach}, H. and {Honkanen}, M.},
        title = "{Two-dimensional Gaussian regression for sub-pixel displacement estimation in particle image velocimetry or particle position estimation in particle tracking velocimetry}",
      journal = {Exp. Fluids},
         year = 2005,
       volume = {38},
        pages = {511-515},
          doi = {10.1007/s00348-005-0942-3},
       adsurl = {https://ui.adsabs.harvard.edu/abs/2005ExFl...38..511N}
}

@article{Ramacciotti2024,
  title = {Simple quantum algorithm to efficiently prepare sparse states},
  author = {Ramacciotti, D. and Lefterovici, A. I. and Rotundo, A. F.},
  journal = {Phys. Rev. A},
  volume = {110},
  issue = {3},
  pages = {032609},
  numpages = {10},
  year = {2024},
  publisher = {American Physical Society},
  doi = {10.1103/PhysRevA.110.032609},
  url = {https://link.aps.org/doi/10.1103/PhysRevA.110.032609}
}

@incollection{Shapiro2003,
    author = {Alexander Shapiro},
    editor = {A. Ruszczynski and A. Shapiro },
    title = {Monte Carlo Sampling Methods},
    booktitle = {Stochastic Programming},
    publisher = {Elsevier},
    year = {2003},
    volume = {10},
    pages = {353--425},
}

@article{Scharnowski_ruleofthumb_2020,
title = {Particle image velocimetry - Classical operating rules from today’s perspective},
journal = {Optics and Lasers in Engineering},
volume = {135},
pages = {106185},
year = {2020},
issn = {0143-8166},
doi = {10.1016/j.optlaseng.2020.106185},
author = {Sven Scharnowski and Christian J. Kähler},
}

@article{Krantz2019,
    author = {Krantz, P. and Kjaergaard, M. and Yan, F. and Orlando, T. P. and Gustavsson, S. and Oliver, W. D.},
    title = {A quantum engineer's guide to superconducting qubits},
    journal = {Applied Physics Reviews},
    volume = {6},
    number = {2},
    pages = {021318},
    year = {2019},
    month = {06},
    issn = {1931-9401},
    doi = {10.1063/1.5089550},
}

@article{Zecchi2025,
doi = {10.1088/2058-9565/addeea},
url = {https://doi.org/10.1088/2058-9565/addeea},
year = {2025},
publisher = {IOP Publishing},
volume = {10},
pages = {035039},
author = {Zecchi, A. A. and Sanavio, C. and Perotto, S. and Succi, S.},
title = {Improved amplitude amplification strategies for the quantum simulation of classical transport problems},
journal = {Quant. Sci. Technol.}
}

@article{Mao2024,
  title = {Toward optimal circuit size for sparse quantum state preparation},
  author = {Mao, R. and Tian, G. and Sun, X.},
  journal = {Phys. Rev. A},
  volume = {110},
  issue = {3},
  pages = {032439},
  numpages = {9},
  year = {2024},
  publisher = {American Physical Society},
  doi = {10.1103/PhysRevA.110.032439},
  url = {https://link.aps.org/doi/10.1103/PhysRevA.110.032439}
}

@article{kahler2016main,
  title={Main results of the 4th {International PIV Challenge}},
  author={K{\"a}hler, C. J. and Astarita, T. and Vlachos, P. P. and Sakakibara, J. and Hain, R. and Discetti, S. and La Foy, R. and Cierpka, C.},
  journal={Exp. Fluids},
  volume={57},
  pages={1--71},
  year={2016},
  publisher={Springer}
}

@article{zuo2024improved,
  title={An improved cross-correlation method for efficient clouds forecasting},
  author={Zuo, H.-M. and Qiu, J. and Li, F.-F.},
  journal={Theor. Appl. Climatol.},
  pages={6491--6505},
volume={155},
  year={2024},
  publisher={Springer}
}

@article{smith1997scientist,
  title={{The Scientist and Engineer’s Guide to Digital Signal Processing}},
  author={Smith, S. W.},
  journal={California Technical Pub},
  year={1997}
}

@article{byerly1965convolution,
  title={Convolution filtering of gravity and magnetic maps},
  author={Byerly, P. E.},
  journal={Geophysics},
  volume={30},
  pages={281--283},
  year={1965},
  publisher={Society of Exploration Geophysicists}
}

@book{Raffel2018,
	address = {Cham},
	edition = {3rd ed.},
	title = {Particle {Image} {Velocimetry}: {A} {Practical} {Guide}},
	shorttitle = {Particle {Image} {Velocimetry}},
	publisher = {Springer International Publishing},
	author = {Raffel, M. and Kähler, C. J. and Kompenhans, J. and Scarano, F. and Wereley, S. T. and Willert, C. E.},
	year = {2018},
}

@misc{Westerweel,
	title = {Particle {Image} {Velocimetry}},
	publisher = {Cambridge University Press},
	author = {Westerweel, J. and Adrian, R.J.},
	year = {2011},
    series    = {Cambridge Aerospace Series},
    isbn      = {9780521440080}
}

@article{lecun1998gradient,
  title={Gradient-based learning applied to document recognition},
  author={LeCun, Y. and Bottou, L. and Bengio, Y. and Haffner, P.},
  journal={Proceedings of the IEEE},
  volume={86},
  pages={2278--2324},
  year={1998},
  publisher={Ieee}
}

@article{hayashi2008understanding,
  title={Understanding the halo-mass and galaxy-mass cross-correlation functions},
  author={Hayashi, E. and White, S. D. M.},
  journal={Mon. Not. R. Astron. Soc.},
  volume={388},
  pages={2--14},
  year={2008},
  publisher={The Royal Astronomical Society}
}

@inproceedings{sarvaiya2009image,
  title={Image registration by template matching using normalized cross-correlation},
  author={Sarvaiya, J. N. and Patnaik, S. and Bombaywala, S.},
  booktitle={International Conference on Advances in Computing, Control, and Telecommunication Technologies},
  pages={819--822},
  year={2009},
  organization={IEEE}
}

@article{farmer1997review,
  title={A review of recent applications of cross-correlation methodologies to human motor unit recording},
  author={Farmer, S. F. and Halliday, D. M. and Conway, B. A. and Stephens, J. A. and Rosenberg, J. R.},
  journal={J. Neurosci. Meth.},
  volume={74},
  pages={175--187},
  year={1997},
  publisher={Elsevier}
}

@Article{Westerweel2004,
  author    = {Westerweel, J. and Geelhoed, P. F. and Lindken, R.},
  journal   = {Exp. Fluids},
  title     = {Single-pixel resolution ensemble correlation for micro-{PIV} applications},
  year      = {2004},
  pages     = {375--384},
  doi       = {10.1007/s00348-004-0826-y},
  publisher = {Springer Science and Business Media LLC},
}

@Article{Scharnowski2011,
  author    = {Scharnowski, S. and Hain, R. and Kähler, C. J.},
  journal   = {Exp. Fluids},
  title     = {Reynolds stress estimation up to single-pixel resolution using {PIV}-measurements},
  year      = {2011},
  pages     = {985--1002},
  volume    = {52},
  doi       = {10.1007/s00348-011-1184-1},
  publisher = {Springer Science and Business Media LLC},
}

@Article{Sciacchitano2012,
  author    = {Sciacchitano, A. and Scarano, F. and Wieneke, B.},
  journal   = {Exp. Fluids},
  title     = {Multi-frame pyramid correlation for time-resolved {PIV}},
  year      = {2012},
  pages     = {1087--1105},
  volume    = {53},
  doi       = {10.1007/s00348-012-1345-x},
  publisher = {Springer Science and Business Media LLC},
}

@Book{Nielsen2010,
  	author = 	 {Nielsen, M. A. and Chuang, I. L.},
  	title = 	 {{Quantum Computation and Quantum Information}},
  	publisher = {Cambridge University Press},
  	year = 	 {2010},
  	address = 	 {Cambridge, UK}
}

@article{Multi_QFT,
      title={{Multidimensional Quantum Fourier Transformation}}, 
      author={P. Pfeffer},
      year={2023},
      journal={arXiv:2301.13835}
}

@article{Ramezani23,
  title = {Quantum multiplication algorithm based on the convolution theorem},
  author = {Ramezani, M. and Nikaeen, M. and Farman, F. and Ashrafi, S. M. and Bahrampour, A.},
  journal = {Phys. Rev. A},
  volume = {108},
  issue = {5},
  pages = {052405},
  numpages = {10},
  year = {2023},
  publisher = {American Physical Society},
  doi = {10.1103/PhysRevA.108.052405},
  url = {https://link.aps.org/doi/10.1103/PhysRevA.108.052405}
}

@article{PRA_Comment,
  title = {{Comment on ``Quantum multiplication algorithm based on the convolution theorem''}},
  author = {Pfeffer, P.},
  journal = {Phys. Rev. A},
  volume = {110},
  issue = {5},
  pages = {056401},
  numpages = {2},
  year = {2024},
  publisher = {American Physical Society},
  doi = {10.1103/PhysRevA.110.056401},
  url = {https://link.aps.org/doi/10.1103/PhysRevA.110.056401}
}

@article{Holmes2023,
  title = {Nonlinear transformations in quantum computation},
  author = {Holmes, Z. and Coble, N. J. and Sornborger, A. T. and Suba{\c{s}}{\i}, Y.},
  journal = {Phys. Rev. Res.},
  volume = {5},
  issue = {1},
  pages = {013105},
  numpages = {20},
  year = {2023},
  publisher = {American Physical Society},
  doi = {10.1103/PhysRevResearch.5.013105},
  url = {https://link.aps.org/doi/10.1103/PhysRevResearch.5.013105}
}

\end{document}